\documentclass[aip,rsi,preprint,graphicx]{revtex4-1} 

\usepackage{graphicx}
\usepackage{times}
\usepackage{nicefrac}
\usepackage{amsmath}
\usepackage{amsfonts}
\usepackage{amssymb}
\usepackage{amsthm}
\usepackage{epsf}
\usepackage{bm}
\usepackage{times}

\usepackage{dcolumn}

\newcolumntype{.}{D{x}{}{-1}}

\draft 

\newcommand{\balpha}{{\mbox{\boldmath$\alpha$}}}

\newcommand{\be}{\begin{eqnarray}}
\newcommand{\ee}{\end{eqnarray}}

\newcommand{\bfr}{{\bf r}}
\newcommand{\bsigma}{{\mbox{\boldmath$\sigma$}}}

\begin{document}


\title{Theory of bound-electron $g$ factor in highly charged ions} 
\thanks{Invited paper published as part of the Proceedings of the Fundamental Constants Meeting,
Eltville, Germany, February, 1-6, 2015.\\}



\author{V.M. Shabaev$^1$, D.A. Glazov$^{1,2,3}$, G. Plunien$^{3}$, and A.V. Volotka$^{1,3}$}
\affiliation{
$^1$Department of Physics, St. Petersburg State University, Ulianovskaya 1, Petrodvorets, St.~Petersburg
198504, Russia \\
$^2$SSC RF ITEP of NRC ``Kurchatov Institute'', Bolshaya Cheremushkinskaya 25, Moscow, 117218, Russia\\
$^3$Institut f\"ur Theoretische Physik, TU Dresden, Mommsenstrasse 13, Dresden, D-01062, Germany
}


\date{\today}

\begin{abstract}
The paper
presents  the current status of the theory of 
bound-electron $g$ factor in highly charged ions. 
The calculations of the relativistic, QED, nuclear  recoil,
nuclear structure, and interelectronic-interaction corrections 
to the $g$ factor are reviewed. 
Special attention is paid to tests of QED effects at strong coupling 
regime and determinations of the fundamental constants. 

\end{abstract}

\pacs{}

\maketitle 

\section{Introduction}
In the past two decades there has been a continuing interest in 
 theoretical studies of the magnetic moments of highly charged ions.
This interest was triggered by the first experiments on  
the $g$ factor of H-like C\cite{haeffner:2000:5308}   and O \cite{verdu:2004:093002} 
and was supported 
by the recent measurements for some higher-$Z$  ions 
\cite{sturm:2011:023002,sturm:2013:R030501,sturm:2014:467}.
The comparison of the theoretical and experimental results has provided
not only the most precise tests of 
quantum electrodynamics (QED) 
with middle-$Z$ ions but has also led to the most precise determination of 
the electron mass in atomic mass units. It is anticipated that in the near future
the measurements of the $g$ factor 
will be extended to heavy ions, including few-electron ions of 
lead and uranium. 
From the theoretical side, to probe the QED effects 
with these measurements, in addition to accurate calculations of various contributions,
one has to find some tricks to avoid the large uncertainties due to
nuclear size and polarization effects. As was shown
 in Refs. \cite{shabaev:2002:062104,shabaev:2006:253002},
this can be done by studying specific differences of the $g$ factors of H-, Li-,
and B-like ions. Therefore, accurate calculations of all these ions are 
required to provide stringent tests of the theoretical methods employed for 
the $g$-factor calculations and to use these studies for   determination of the fundamental
constants. The present paper contains an overview of 
these calculations. 

The relativistic units ($\hbar=c=m =1$) are used throughout the paper.


\section{The $g$ factor of H-like ions}

The $g$ factor of an ion can be defined as the ratio of the magnetic
moment of the ion to its mechanical moment expressed in the Bohr's magnetons.
Alternatively, it can be defined as the dimensionless coefficient in the
linear part of the Zeeman splitting:
\begin{eqnarray}
\Delta E=g\mu_{0}{\cal H}M_J\,,
\end{eqnarray}
where $\mu_{0}=|e|\hbar/(2mc)$ is the Bohr magneton and $M_J$ is the 
angular momentum projection on the direction of the homogeneous magnetic
field ${\cal H}$. The total theoretical value of the $g$ factor of a hydrogenlike ion
is given by a sum
\begin{eqnarray}
 g=g_{\rm D}+\Delta g_{\rm QED}+\Delta g_{\rm NR}+\Delta g_{\rm NS}+\Delta g_{\rm NP}\,,
\end{eqnarray}
where $g_{\rm D}$ is the point-nucleus Dirac value, $\Delta g_{\rm QED}$
is the QED correction, and the last three terms denote the nuclear recoil (NR),
nuclear size (NS), and nuclear polarization (NP) corrections, respectively.
The point-nucleus Dirac value  can easily 
be evaluated analytically:
\begin{eqnarray}  \label{dir}
g_{\rm D} = \frac{\kappa}{2j(j+1)}\Bigl(\frac{2\kappa E}{m}-1\Bigr)\,,
\end{eqnarray}
where  $E$ is the Dirac energy for the point-charge nucleus
and $\kappa=(-1)^{j+l+1/2}(j+1/2)$ is the angular-momentum-parity
quantum number.
For an $ns$ state it yields  \cite{breit:1928:649}
\begin{eqnarray} \label{dir_ns}
g_{\rm D}&=&{ 2}+\frac{4}{3}\frac{E-m}{m}\nonumber\\
&=&{2}-\frac{2}{3}\frac{(\alpha Z)^2}{n^2}
+\Bigl(\frac{1}{2n}-\frac{2}{3}\Bigr)\frac{(\alpha Z)^4}{n^3}
+\cdots\,,
\end{eqnarray}
where the first term, 2, corresponds to the free-electron Dirac $g$ factor
and the other terms are the relativistic binding corrections.  
The QED correction can also be represented as a sum of the free-electron
QED contribution and the binding-QED correction:
\begin{eqnarray}
\Delta g_{\rm QED}={\Delta g_{\rm free-QED}}+\Delta g_{\rm bind.-QED}\,.\nonumber
\end{eqnarray}
The free-electron $g$ factor is given by
\begin{eqnarray}
g_{\rm free}&=&2+\Delta g_{\rm free-QED}+\Delta\nonumber\\
&=&2+2\Bigl[A^{(2)}\frac{\alpha}{\pi}+  A^{(4)}\Bigl(\frac{\alpha}{\pi}\Bigr)^2
+ A^{(6)}\Bigl(\frac{\alpha}{\pi}\Bigr)^3
+\cdots \Bigr] +\Delta\,,
\end{eqnarray}
where the $\Delta$ term denotes the sum of the hadronic and weak
contributions. To date, the $A^{(2n)}$ coefficients have been calculated up to
n=5 (see Refs. \cite{aoyama:2012:111807,kinoshita:2014:1430003} and references therein). With these calculations,
the theoretical
accuracy of the free-electron $g$ factor is presently limited by the accuracy
of the fine structure constant $\alpha$. As the result, the comparison of the theory
and the corresponding free-electron $g$-factor
 experiment \cite{hanneke:2014:120801} has provided 
the most precise determination of $\alpha$ \cite{aoyama:2012:111807,kinoshita:2014:1430003}. 

\subsection{Binding-QED corrections}
For an $ns$ state, the binding-QED correction to the
lowest order in $\alpha Z$ and to all orders in $\alpha$ is given by
\cite{grotch:1970:1605,faustov:1970:422,close:1971:400,eides:1997:191,czarnecki:2000:012509,karshenboim:2001:651}:
\begin{eqnarray} \label{qed_lo}
 \Delta g_{\rm bind.-QED}^{\rm (l.o.)}=
\Delta g_{\rm free-QED}\frac{(\alpha Z)^2}{6n^2}\,,
\end{eqnarray}
where  $\Delta g_{\rm free-QED}$
is the total free-electron QED correction. The formula (\ref{qed_lo}) can be derived
using the Pauli operator
\begin{eqnarray}
H_{\rm rad}=\frac{|e|}{2m}
\frac{g_{\rm free}-2}{2}[\beta (\bsigma\cdot{\bf{\cal H}})-
i\beta(\balpha\cdot{\bf{\cal E}})]\,,
\end{eqnarray}
where ${\bf {\cal E}}=|e|Z\bfr/(4\pi r^3)$,
$\balpha$ and $\beta$ are the Dirac matrices.
The expectation value of $H_{\rm rad}$ is evaluated
with the Dirac wave function of the electron that accounts for
the interaction with the homogeneous magnetic field
to first order in ${\cal H}$.  The first-order correction
 to the electronic wave function
due to the interaction with the magnetic field 
is easily obtained 
using the generalized virial relations for the
Dirac equation \cite{shabaev:1991:4479,shabaev:2003:97}. 

\begin{figure}
\includegraphics{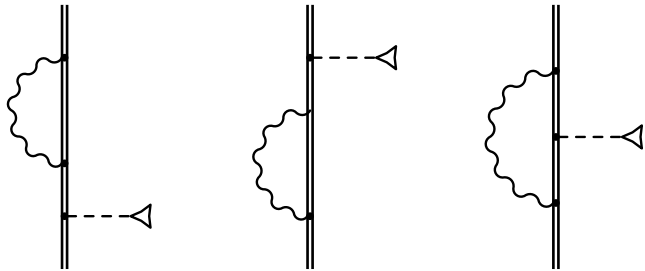}
\caption{\label{g-se} The self-energy corrections to the bound-electron $g$ factor.}%
\end{figure}

\begin{figure}
\includegraphics{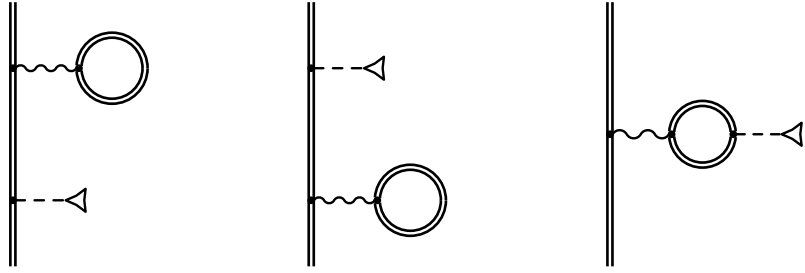}
\caption{\label{g-vp} The vacuum-polarization corrections to the bound-electron $g$ factor.
}%
\end{figure}

The evaluation of the QED correction
to all orders in the $\alpha Z$ parameter is a much more difficult problem.
This correction consists of two contributions: the self-energy (SE) and
vacuum-polarization (VP) ones. The corresponding Feynman diagrams are 
given in Figs. \ref{g-se} and \ref{g-vp}, respectively,
where the dashed line terminated by a triangle represents the interaction 
with the external magnetic field.
For  the ground $1s$ state
the SE correction to all orders in
$\alpha Z$ was first evaluated in Refs. \cite{blundell:1997:1857,persson:1997:R2499,beier:2000:032510,beier:2000:79}.
However, the first experiments on the $g$ factor of H-like C\cite{haeffner:2000:5308} and O
\cite{verdu:2004:093002} demanded
 more precise calculations of this correction. 
The required accuracy was achieved 
in Refs. \cite{yerokhin:2002:143001,yerokhin:2004:052503}
and further improved 
in Refs. \cite{yerokhin:2008:163001,yerokhin:2010:012502}. 

The VP contributions are generally represented by a sum 
of the Uehling and the Wichmann-Kroll terms. 
The Uehling 
term is defined by the lowest-order contributions in
the expansion of the
fermion loops in powers of the electron-nucleus interaction,
that are not ruled out by the Furry theorem.
In the Uehling approximation only the electric-loop
diagrams contribute to the $g$ factor value. 
These are
the first two diagrams in Fig. \ref{g-vp}.
As to the magnetic-loop VP
contribution (the third diagram in Fig. \ref{g-vp}), it vanishes in the Uehling approximation.
The evaluation of
the Uehling and Wichmann-Kroll
 corrections to all orders in  $\alpha Z$ 
was considered in Refs. \cite{persson:1997:R2499,beier:2000:032510,beier:2000:79}.
 For the point-charge
nucleus, the Uehling contribution can be evaluated 
analytically \cite{karshenboim:2001:81}.
The evaluation of the magnetic-loop Wichmann-Kroll part  
to the lowest order in  $\alpha Z$ for an $ns$ state 
gives \cite{karshenboim:2002:321}
\be
\Delta g_{\rm VP}^{\rm magn}=\frac{7}{216}\frac{\alpha(\alpha Z)^5}{n^3}\,.
\ee
The next-to-leading contribution of magnetic loop was calculated in Ref.
\cite{lee:2005:052501}.

The two-loop QED correction to the order $\alpha^2(\alpha Z)^2$ 
for an $ns$ state is determined by Eq. (\ref{qed_lo}). The evaluation
of this correction to the order  $\alpha^2(\alpha Z)^4$, performed in
Refs. \cite{pachucki:2004:150401,pachucki:2005:022108}, yields
\begin{eqnarray}
\Delta g_{\rm two-loop}^{\rm (h.o.)}
&=&\Bigl(\frac{\alpha}{\pi}\Bigr)^2
\frac{(\alpha Z)^4}{n^3}
\Bigl\{\frac{28}{9}{\rm ln}[(\alpha Z)^{-2}]
+\frac{258917}{19440}
 -\frac{4}{9}{\rm ln}k_0\nonumber\\
&& -\frac{8}{3}{\rm ln}k_3 +\frac{113}{810}\pi^2-\frac{379}{90}\pi^2 {\rm ln}2
+\frac{379}{60}\zeta(3)\nonumber\\
&& +\frac{1}{n}\Bigl[-\frac{985}{1728} -\frac{5}{144}\pi^2
 +\frac{5}{24}\pi^2 {\rm ln}2-\frac{5}{16}\zeta(3)
\Bigr]\Bigr\}\,,
\end{eqnarray}
where $\zeta(s)$ is the Rieman zeta function,
${\rm ln}k_0(1s)=2.984128556$, ${\rm ln}k_0(2s)=2.811769893$,
${\rm ln}k_3(1s)=3.272806545$, and ${\rm ln}k_3(2s)=3.546018666$.
Finally, we note that the calculations of two-loop QED corrections with
the closed fermion loops were considered
for the $1s$ state in Refs. \cite{jentschura:2009:044501,yerokhin:2013:042502}.

\subsection{Nuclear recoil corrections}

It is known  \cite{bethe} that  
for an $ns$ state  
the nuclear recoil correction
to the $g$ factor
 is of pure relativistic origin. This means that,
in contrast to the $p$ states,
its expansion in the parameter $\alpha Z$ starts with 
the  $(\alpha Z)^2$ terms, which can be derived from the 
Breit equation. An extention of this equation, including 
the interaction due to the anomalous magnetic moment of 
free electron, allows also to account for the lowest-order 
radiative recoil effect. As the result,
the nuclear recoil correction to the orders up to
  $(\alpha Z)^2$,  $(\alpha/\pi)$,
and $(m/M)^2$  is given by
\cite{faustov:1970:422,grotch:1971:59,close:1971:400,eides:1997:191,pachucki:2008:012504,eides:2010:100402}
\begin{eqnarray}
\Delta g_{\rm NR}^{\rm(l.o.)}&=&\frac{(\alpha Z)^2}{n^2}
\Bigl[\frac{m}{M}-(1+Z)\Bigl(\frac{m}{M}\Bigr)^2\Bigr]\nonumber\\
&&+ \frac{\alpha}{\pi}
\frac{(\alpha Z)^2}{n^2}
\Bigl[-\frac{1}{3}\,\frac{m}{M}+\frac{3-2Z}{6}
\Bigl(\frac{m}{M}\Bigr)^2\Bigr]\,,
\end{eqnarray}
where $M$ is the nuclear mass.
It should be noted that this correction does not depend on the nuclear spin
\cite{close:1971:400,eides:1997:191,pachucki:2008:012504,eides:2010:100402}.

To evaluate the nuclear recoil correction for high-$Z$ ions, we need first to derive 
the exact $\alpha Z$-dependence formula for the recoil effect, at least to the first order
in $m/M$. The desired formula was derived in Ref. \cite{shabaev:2001:052104}. According to this 
formula, which was also confirmed in Ref. \cite{yelkhovsky:2001:arxiv}, the recoil correction 
to the $g$ factor to the first order in $m/M$ and to all orders in  $\alpha Z$
is given by  ($e <0$)
\begin{eqnarray}
\Delta g_{\rm NR}&=&\frac{1}{\mu_0 m_a}\frac{i}{2\pi M}
\int_{-\infty}^{\infty} d\omega\;
\Biggl[\frac{\partial}{\partial {\cal H}}
\langle \tilde{a}|[p^k-D^k(\omega)+eA_{\rm cl}^k]
\nonumber\\
&&\times\tilde{G}(\omega+\tilde{E}_a)
[p^k-D^k(\omega)+eA_{\rm cl}^k]
|\tilde{a}\rangle
\Biggr]_{{\cal H}=0}\,.
\label{06recoilt}
\end{eqnarray}
Here $m_a$ is the angular momentum
projection of the state under consideration, 
${\bf A}_{\rm cl}=[\mbox{\boldmath ${\cal H}$}
\times {\bf r}]/2$ is the 
vector potential of the homogeneous
magnetic field $\mbox{\boldmath ${\cal H}$}$ 
directed along the $z$ axis,
 $p^k=-i\nabla^k$ is the momentum operator, 
$D^k(\omega)=-4\pi\alpha Z\alpha^l D^{lk}(\omega)$,
\begin{eqnarray} \label{06photon}
D^{il}(\omega,{\bf r})&=&-\frac{1}{4\pi}\Bigl\{\frac
{\exp{(i|\omega|r)}}{r}\delta_{il}+\nabla^{i}\nabla^{l}
\frac{(\exp{(i|\omega|r)}
-1)}{\omega^{2}r}\Bigr\}\,
\end{eqnarray}
is the transverse part of the photon propagator in the Coulomb 
gauge. All the quantities marked with a tilde  
(the wave function, the energy, and the Coulomb Green function
$\tilde{G}(\omega)=\sum_{\tilde{n}}|\tilde{n}\rangle \langle 
\tilde{n}|[\omega-\tilde{E}_n(1-i0)]^{-1}$)
are assumed to be calculated in presence of the
magnetic field $\mbox{\boldmath ${\cal H}$}$.
For the practical calculations, it is convenient to represent
the expression (\ref{06recoilt}) as a sum 
of a lower-order term and a higher-order
term, $\Delta g_{\rm NR}=\Delta g_{\rm NR}^{\rm (l.o.)}
+\Delta g_{\rm NR}^{\rm (h.o.)}$, where 
\begin{eqnarray} \label{06recoil1}
\Delta g_{\rm NR}^{\rm (l.o.)}
&=&\frac{1}{\mu_0 m_a}\frac{1}{2M} 
\Biggl[\frac{\partial}{\partial {\cal H}}
\langle \tilde{a}|\Bigl\{
{\bf p}^2-\frac{\alpha Z}{r}[(\mbox{\boldmath $\alpha$}
\cdot {\bf p})
+(\mbox{\boldmath $\alpha$}\cdot {\bf n})
({\bf n}\cdot{\bf p})]\Bigr\}|\tilde{a}\rangle
 \Biggr]_{{\cal H}=0}\nonumber \\
&&-\frac{1}{m_a}
\frac{m}{M}\langle a|\Bigl([{\bf r}\times {\bf p}]_z
-\frac{\alpha Z}{2r}[{\bf r}\times \mbox{\boldmath $\alpha$}]_z
\Bigr)|a\rangle\,,\\
\Delta g_{\rm NR}^{\rm (h.o.)}
&=&\frac{1}{\mu_0 m_a}\frac{i}{2\pi M}
\int_{-\infty}^{\infty} d\omega\;
\Biggl[\frac{\partial}{\partial {\cal H}}
\langle \tilde{a}|\Bigl(D^k(\omega)-\frac{[p^k,V]}{\omega+i0}\Bigr)
\nonumber\\
&&\times\tilde{G}(\omega+\tilde{E}_a)
\Bigl(D^k(\omega)+\frac{[p^k,V]}{\omega+i0}\Bigr)|\tilde{a}\rangle
\Biggr]_{{\cal H}=0}\,,
\label{06recoil2}
\end{eqnarray}
where  $V(r)=-\alpha Z/r$ is the Coulomb
potential of the nucleus and
${\bf n}={\bf r}/r$.
The low-order term can be evaluated analytically employing the 
 generalized virial relations \cite{shabaev:1991:4479,shabaev:2003:97}. This
yields \cite{shabaev:2001:052104}
\begin{eqnarray} \label{06shabaeveq112}
\Delta g_{\rm NR}^{\rm (l.o.)}=-\frac{m}{M}\,
\frac{2\kappa^2E^2+\kappa m E-m^2}{2m^2j(j+1)}\,,
\end{eqnarray}
where $E$ is the Dirac energy.
{To the two lowest orders in {$\alpha Z$}, we get}
\begin{eqnarray} \label{06shabaeveq113}
\Delta g_{\rm NR}^{\rm (l.o.)}
=-\frac{m}{M}\,
\frac{1}{j(j+1)}\Bigl[\kappa^2+\frac{\kappa}{2}
-\frac{1}{2}-\Bigl(\kappa^2+\frac{\kappa}{4}\Bigr)
\frac{(\alpha Z)^2}{n^2}\Bigr]\,.
\end{eqnarray}
{For the { $1s$} state, the exact formula (\ref{06shabaeveq112})
 takes the form:}
\begin{eqnarray} 
\Delta g_{\rm NR}^{\rm (l.o.)}=\frac{m}{M}(\alpha Z)^2-\frac{m}{M}
\frac{(\alpha Z)^4}{3[1+\sqrt{1-(\alpha Z)^2}]^2}\,.\nonumber
\end{eqnarray}
{The higher-order term can be represented as}
\begin{eqnarray}
\Delta g_{\rm NR}^{\rm (h.o.)}=\frac{m}{M}\,\frac{(\alpha Z)^5}{n^3}P(\alpha Z)\,.\nonumber
\end{eqnarray}
The numerical evaluation of the function {$P(\alpha Z)$} for the { $1s$}  state
was performed in Ref. \cite{shabaev:2002:091801}.

\subsection{Nuclear size and polarization corrections}
The finite nuclear size correction to the $g$ factor can be calculated numerically (see, e.g. 
Refs. \cite{beier:2000:032510,beier:2000:79}). The perturbative evaluation of this correction  
to two lowest orders in $\alpha Z$ yields
for an $ns$ state \cite{glazov:2002:408}:
\begin{eqnarray}
  \Delta g_{\rm NS}
&=&\frac{8}{3n^3} (\alpha Z)^4 m^2
\langle r^2 \rangle_{\rm nuc}
  \Biggl[ 1+(\alpha Z)^2
  \Biggl( \frac{1}{4} + \frac{12n^2-n-9}{4n^2(n+1)}
\nonumber\\
&& + 2\psi(3) -\psi(2+n)
   - \frac{\langle r^2 \ln(2\alpha Zmr/n) \rangle_{\rm nuc}}
  {\langle r^2 \rangle_{\rm nuc}} \Biggr) \Biggr]\,,
\label{06shabaevgl2}
\end{eqnarray}
where $\psi(x)=\frac{d}{dx}\ln\Gamma(x)$.
The first term in the right-hand side of the equation (\ref{06shabaevgl2})
was first derived in Ref. \cite{karshenboim:2000:380}. 
An approximate formula which expresses the nuclear size correction to the $g$ factor
in terms of the corresponding correction to the binding energy was derived in Ref. \cite{karshenboim:2005:042101}.
The dependence 
of the nuclear size correction on the nuclear deformation parameters was 
studied in Ref. \cite{zatorski:2012:063005}.

\begin{figure}
\includegraphics{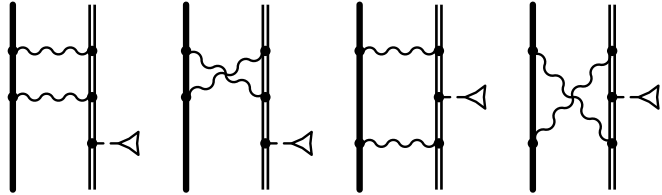}
\caption{\label{g-np} The nuclear-polarization corrections to the bound-electron $g$ factor.}%
\end{figure}

The nuclear polarization correction to the $g$ factor is defined by 
the Feynman 
diagrams presented in Fig. \ref{g-np}. In these diagrams the double, thick, and wavy lines
correspond to the electron, nucleus, and photon propagators, respectively. 
The dashed line ended by a triangle 
denotes the interaction of electron with the external magnetic field. 
The evaluations of these diagrams for some middle- and high-$Z$ ions
were performed in Refs. \cite{nefiodov:2002:081802,volotka:2014:023002}. 
In addition to the
nuclear polarization contribution, there exists also a nuclear magnetic
susceptibility correction, which is defined by the one-photon exchange
diagrams with the magnetic interaction attached to the nuclear line.
The calculations performed in Refs. \cite{jentschura:2006:102,volotka:2014:023002}
showed that this correction is rather small.

\subsection{Comparison of theory and experiment: determination of the electron mass}
First high-precision measurement of the $g$ factor with highly charged ions
was accomplished for $^{12}$C$^{5+}$ in Ref. \cite{haeffner:2000:5308}. The experiment was performed
using a single ion confined in a Penning ion trap. The experimental value
of the $g$ factor was presented as
\be
 g_{\rm exp}=2(q/|e|)(m/M_{\rm ion})(\omega_L/\omega_c)\,,
\ee
where  $\omega_c=(q/M_{\rm ion}) {\cal H}$
is the cyclotron frequency,  
{$\omega_L=\Delta E/\hbar$}
is the Larmor precession frequency, 
{$M_{\rm ion}$} is the ion mass, 
and {$q$} is the ion charge. 
The experimental accuracy of the $\omega_L/\omega_c$ ratio was so high that 
the uncertainty of  $g_{\rm exp}$ was mainly due to the uncertainty of the
value of the electron mass. 
This stimulated 
high-precision calculations of the nuclear recoil 
and one-loop QED corrections \cite{shabaev:2001:052104,shabaev:2002:091801,yerokhin:2002:143001}. As the result,
the theoretical accuracy was significantly improved, and  the comparison
of the theory and the experiment led to a four-times improvement of 
the accuracy of the electron mass  \cite{beier:2002:011603,yerokhin:2002:143001,mohr:2012:1527}. 
Later \cite{verdu:2004:093002}, 
the $g$ factor
of  $^{16}$O$^{7+}$ was measured to a similar accuracy.
The value of the electron mass derived from the comparison of this 
experiment with the related theory agreed with the
determination on $^{12}$C$^{5+}$.

In Refs. \cite{sturm:2011:023002,sturm:2013:R030501} the $g$ factor of  $^{28}$Si$^{13+}$
was measured to amount  $g_{\rm exp}=1.995 348 959 10(7)(7)(80)$,
where the first and second errors represent 
the statistical and systematic uncertainties, while 
the third one is due to the current uncertainty of the electron mass.
The theoretical contributions to the $g$ factor of  $^{28}$Si$^{13+}$
are presented in Table \ref{H-like-Si}.
These experiment and theory provide the most accurate to-date 
test of bound-state QED with middle-$Z$ ions.
The current theoretical uncertainty, which
is two times bigger than the experimental one, 
 is mainly 
defined by uncalculated two-loop QED corrections of order $\alpha^2(\alpha Z)^5$
and higher. This uncertainty can be reduced in a combination of the 
corresponding theoretical and experimental values for two different H-like ions.
This idea was explored in Ref. \cite{sturm:2014:467}, where the high-precision measurements
of the $g$ factors of $^{12}$C$^{5+}$ and  $^{28}$Si$^{13+}$ were combined with
the theory to extract a new value of the electron mass which is by a factor of 13 
more precise than the previously accepted value. Namely, the experimental value
of the $g$ factor, which is determined by 
\be \label{nat1}
g_{\rm exp}=2(Z-1)(m/M_{\rm ion}) 
(\omega_L/\omega_c)\,,
\ee
was fitted as
\be \label{nat2}
 g_{\rm exp}= g_{\rm theor}^* +({\alpha}/{\pi})^2
(\alpha Z)^5 b_{50}\,,
\ee
where $g_{\rm theor}^*$ is the theoretical value which incorporates
only the known contributions. 
Then, the equations (\ref{nat1})-(\ref{nat2}) give
\begin{eqnarray}
\Bigl(\frac{\alpha}{\pi}\Bigr)^2(6\alpha)^5b_{50}
&=&2(6-1)\frac{m}{M_{^{12}{\rm C}^{5+}}} 
\Bigl(\frac{\omega_L}{\omega_c}\Bigr)_{^{12}{\rm C}^{5+}}-g_{\rm theor}^*[^{12}{\rm C}^{5+}]\,,\\
\Bigl(\frac{\alpha}{\pi}\Bigr)^2(14\alpha)^5b_{50}
&=&2(14-1)\frac{m}{M_{^{28}{\rm Si}^{13+}}} 
\Bigl(\frac{\omega_L}{\omega_c}\Bigr)_{^{28}{\rm Si}^{13+}}-g_{\rm theor}^*[^{28}{\rm Si}^{13+}]\,.
\end{eqnarray}
The solution of these equations yields \cite{sturm:2014:467}
 $m=0.000 548 579 909 067(14)(9)(2)$ u and
$b_{50}=-4.0(5.1).$ The indicated uncertainties  include also an uncertainty
due to the omitted higher-order QED corrections, 
$\sim ({\alpha}/{\pi})^2 (\alpha Z)^6{\rm ln}^k[(\alpha Z)^{-2}]$,
in Eq. (\ref{nat2}).

\section{The $g$ factor of Li-like ions}
The theoretical accuracy of the $g$ factor for high-$Z$ ions is strongly limited
by the uncertainties of the nuclear effects. In particular, it makes impossible 
tests of the two-loop QED contributions by the direct comparison of the theory and experiment
on the $g$ factor of H-like lead or uranium ions (see, e.g., Refs. \cite{shabaev:2002:062104,volotka:2014:023002}).
To extend the region accessible to the QED tests with the $g$-factor experiments, 
it was proposed \cite{shabaev:2002:062104}
 to study a specific difference of the $g$ factors of H- and Li-like ions
of the same isotope:
\be \label{dif1}
g'=g_{(1s)^22s}-\xi g_{1s}\,,
\ee
where $g_{1s}$ and  $g_{(1s)^22s}$ are the ground-state $g$ factors of H- and
Li-like ions, respectively, and the parameter
$\xi$ is chosen to cancel the nuclear size effect.
It can be shown \cite{shabaev:2002:062104} 
that both the parameter $\xi$ and the difference $g'$  are very
stable with respect to variations of the nuclear structure parameters.
As the result, this difference can be calculated to a much higher accuracy
than each of the $g$ factors. This stimulated significant efforts in calculations
of the $g$ factor of Li-like ions
\cite{glazov:2004:062104,glazov:2006:330,volotka:2009:033005,glazov:2010:062112,volotka:2013:636,volotka:2014:253004}.

The theoretical value of the $g$ factor of a Li-like ion can be written as
\begin{eqnarray}\label{g-li}
 g=g_{\rm one-elec}+\Delta g_{\rm int}+\Delta g_{\rm scr.-QED}\,,
\end{eqnarray}
where $g_{\rm one-elec}$ incorporates all the one-electron contributions,
which were considered in the previous section, $\Delta g_{\rm int}$ is 
the interelectronic-interaction contribution, and  $\Delta g_{\rm scr.-QED}$
denotes the screened QED correction. 
The interelectronic-interaction and screened QED corrections are evaluated
using the perturbation theory in the parameter {$1/Z$}. 

\begin{figure}
\includegraphics{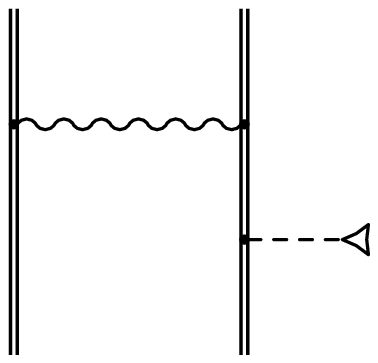}
\caption{\label{g-1ph} The one-photon-exchange correction to the bound-electron $g$ factor.}%
\end{figure}

\begin{figure}
\includegraphics{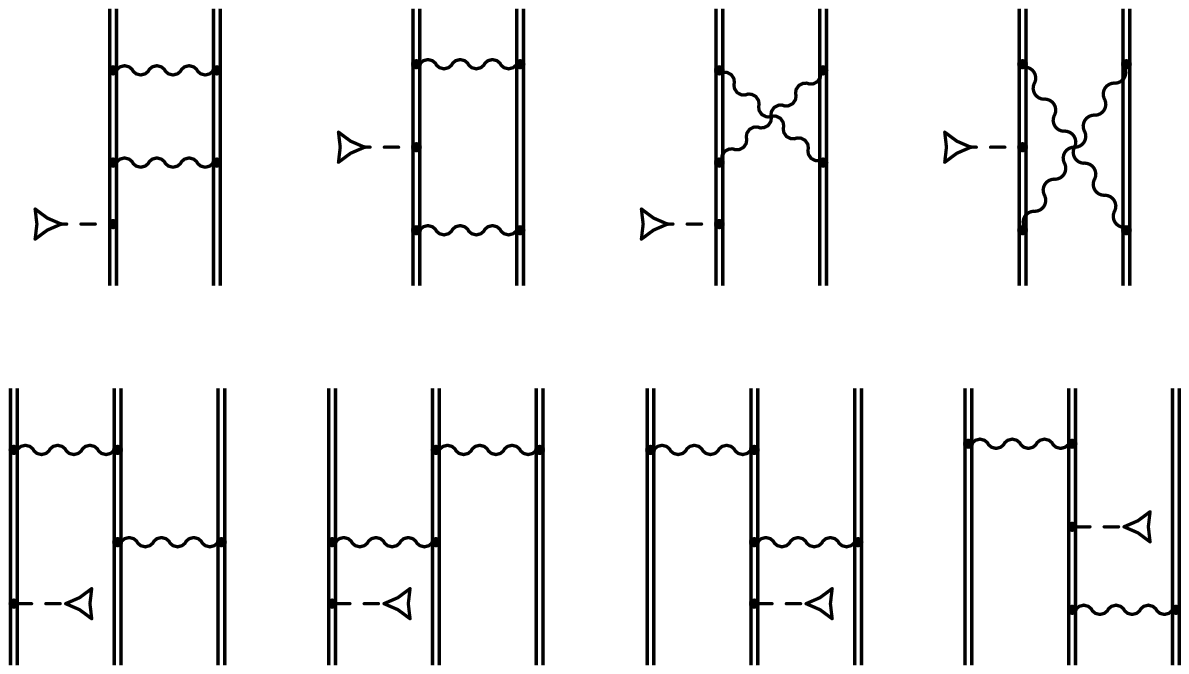}
\caption{\label{g-2ph} The two-photon-exchange corrections to the bound-electron $g$ factor.}%
\end{figure}

The interelectronic-interaction
contribution can be represented as
\be
\Delta g_{\rm int} = \Delta g_{\rm int}^{(1)}+ \Delta g_{\rm int}^{(2)}+\Delta g_{\rm int}^{(3+)}\,,
\ee
where the terms 
\be \label{b1}
\Delta g_{\rm int}^{(1)} =\frac{1}{Z}(\alpha Z)^2 B(\alpha Z)\,,\\
\Delta g_{\rm int}^{(2)} =\frac{1}{Z^2}(\alpha Z)^2 C(\alpha Z)\,
\label{c2}
\ee
denote the contributions of the first and second orders in $1/Z$, which 
are defined by the Feynman
diagrams depicted in Figs. \ref{g-1ph} and \ref{g-2ph}, respectively. 
The term $\Delta g_{\rm int}^{(3+)}$ includes all the interelectronic-interaction
corrections of the third and higher orders in $1/Z$. 
In formulas (\ref{b1}) and  (\ref{c2})
 the factor $(\alpha Z)^2$ accounts for 
 the relativistic origin of  the interelectronic-interaction effects
on the $g$ factor. The function $B(\alpha Z)$ 
 was first evaluated in Ref. \cite{shabaev:2002:062104}.  
The exact calculation of the  function  
$C(\alpha Z)$ is a much more difficult task. This calculation within the rigorous QED
approach was performed in Ref. \cite{volotka:2014:253004}. 
In addition, in  that paper the contribution
$\Delta g_{\rm int}^{(3+)}$ was evaluated within the Breit approximation
using the large-scale configuration-interaction Dirac-Fock-Sturm method 
\cite{tupitsyn:2003:022511,glazov:2004:062104,tupitsyn:2005:062503,shabaev:2005:062105}. 
To accelerate the convergence of the $1/Z$ expansion,
the calculations were performed starting with an effective spherically symmetric
potential which partly accounts for the electron-electron interaction effects. To avoid 
the double counting of these effects, the related subtraction was carried out
in the higher orders of the perturbation theory. 

\begin{figure}
\includegraphics{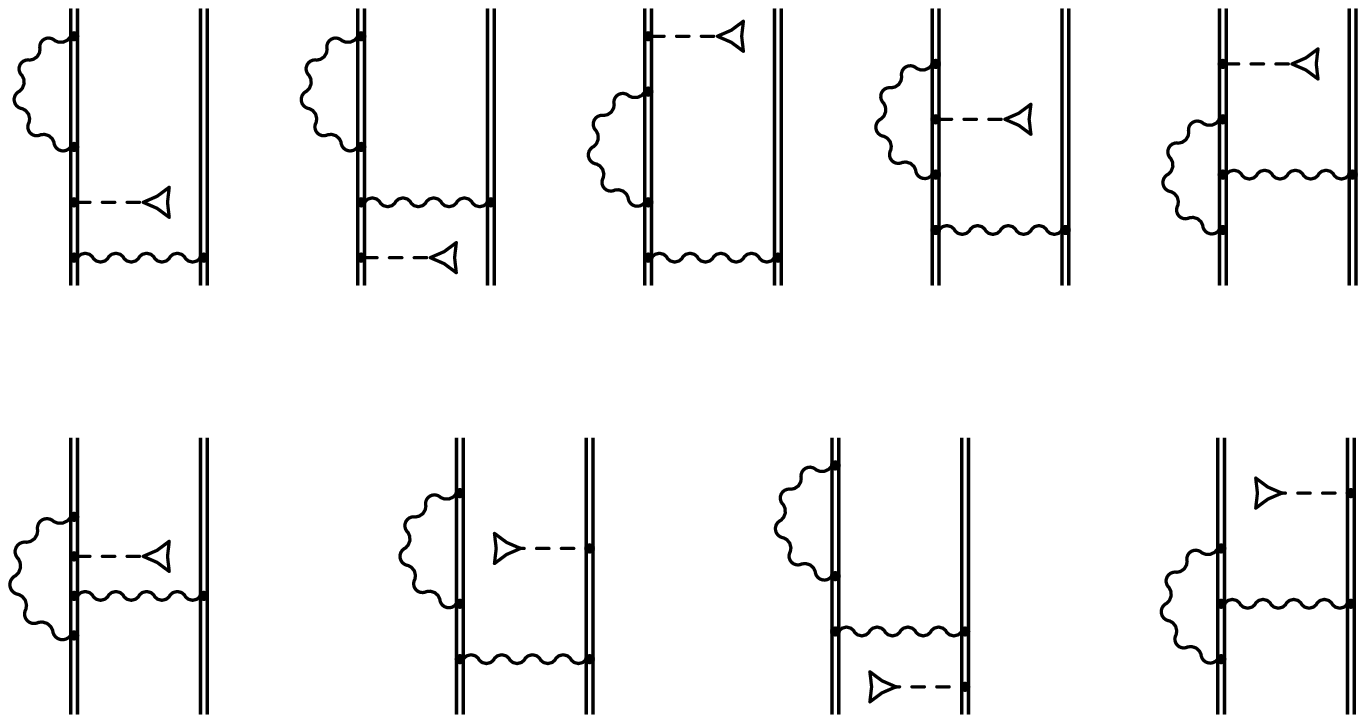}
\caption{\label{g-se-1ph}  The screened SE corrections to the bound-electron $g$ factor.}%
\end{figure}

\begin{figure}
\includegraphics{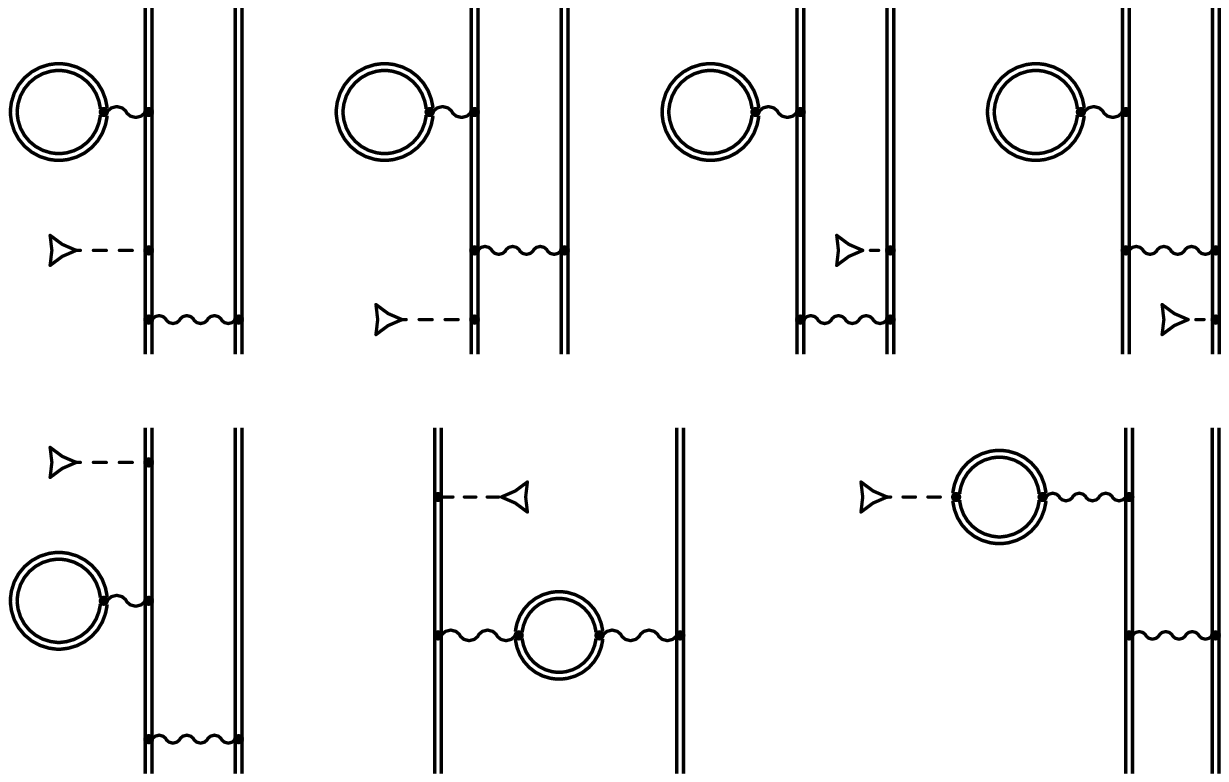}
\caption{\label{g-vp-1ph}  The screened VP corrections to the bound-electron $g$ factor.}%
\end{figure}

The calculation of the one-loop QED corrections to the $g$ factor of Li-like ions
with an effective local potential, which partly accounts for the screening effect, 
was first performed in Ref. \cite{glazov:2006:330}.
The screened SE corrections and the Uehling part of the 
screened VP corrections, presented in
 Figs. \ref{g-se-1ph} and  \ref{g-vp-1ph}, respectively, were evaluated 
in Refs. \cite{volotka:2009:033005,glazov:2010:062112,volotka:2014:253004}. 
As in the case of the two-photon exchange diagrams,
the calculations were accomplished starting with an effective local potential.

In Table \ref{Li-like-Si} the theoretical prediction
for  the $g$ factor of  $^{28}$Si$^{11+}$ is compared with the 
recent experiment \cite{wagner:2013:033003}. To date,
these experiment and theory provide the most accurate 
test of many-electron QED effects with middle-$Z$ ions.

\section{Future prospects}
\subsection{Tests of QED and determination of the nuclear charge radii}
The measurements of 
the $g$ factors of heavy few-electron ions,
that are anticipated in the near future at the HITRAP facilities
in Darmstadt and at MPIK in Heidelberg,
will provide the most  precise tests of the magnetic
sector of QED at strong electric field. Moreover,  the study
of the $g$ factor 
can give a unique opportunity to test QED effects  with
highly charged ions
beyond the
external field approximation (the Furry picture of QED). 
The most appropriate way to access bound-state QED
beyond the Furry picture would consist in studying the isotope shifts
of the $g$ factor of highly charged ions. In Table \ref{g-ca}
we present the theoretical contributions to the isotope shift 
of H-like calcium. The nuclear size effect was evaluated with 
the nuclear charge radii taken from Ref. \cite{angeli:2013:69}.
The uncertainty of this correction includes both the nuclear radius and shape variation
effects.
As one can see from the table,
the current theoretical uncertainty of the  $^{40}{\rm Ca}^{19+} - \; ^{48}{\rm Ca}^{19+}$
isotope shift 
is about 9\% of the QED nuclear recoil contribution.
This would allow tests
of bound-state QED
beyond the external field approximation with highly charged ions,
provided the corresponding shift is measured to the required accuracy.
Alternatively, the study of the isotope shift can provide a
determination of 
the nuclear-charge-radius difference to a high
precision.

\subsection{Determination of the nuclear magnetic moments}
The extention of the $g$ factor measurements to ions with non-zero nuclear spin
would give an access to the nuclear magnetic moments \cite{werth:2001:204}.
To the lowest order, the $g$ factor of an H-like ion with a nonzero 
nuclear spin is given by (see, e.g., Ref. \cite{bethe}):  
\begin{eqnarray} 
\label{g-ion}
g_{\rm ion} &= &g_{\rm D}\frac{F(F+1)+j(j+1)-I(I+1)}{2F(F+1)}\nonumber\\
&&- \frac{m}{m_p} g_{\rm N}\frac{F(F+1)+I(I+1)-j(j+1)}{2F(F+1)}\,,
\end{eqnarray}
where $F$ is the total angular momentum of the ion, $j$ and $I$  are the 
electronic and nuclear angular momenta, respectively, $g_{\rm D}$ is
the Dirac value of the electronic $g$ factor given by Eq. (\ref{dir}),  $ g_{\rm N} $
is the nuclear $g$ factor, and $m_p$ is the proton mass. 
Various corrections to Eq. (\ref{g-ion}) 
for H- and Li-like ions as well as the corresponding corrections
to the Breit-Rabi formula, which describes the Zeeman effect
on the hyperfine-structure levels, were derived in Refs. 
\cite{moskovkin:2004:032105,moskovkin:2006:052506,moskovkin:2008:637,moskovkin:2008:063421,yerokhin:2011:043004,yerokhin:2012:022512}. 
As was discussed in detail in Ref. \cite{quint:2008:032517}, 
due to the absence of an uncertainty caused by the diamagnetic shielding,  
these studies can provide a determination of the nuclear magnetic moment with 
unprecedented accuracy.

\subsection{Access to the nonlinear Zeeman effect}
The laser-microwave double-resonance technique 
\cite{quint:2008:032517,vogel:2010:1,lindenfels:2011:79,lindenfels:2013:023412,vogel:2013:505} 
allows precise measurements of the Zeeman splittings of fine- and 
hyperfine-structure levels in a Penning trap. This,  apart from the linear Zeeman effect,
provides an access to the 
second- and third-order Zeeman effect with highly charged ions. The influence of the
higher-order Zeeman
effects on the fine-structure levels in boronlike argon was studied
in Refs. \cite{lindenfels:2013:023412,glazov:2013:014014}.
The related experiments 
are currently under preparation at GSI (Darmstadt).

\subsection{Determination of the fine structure constant}
Finally, let us discuss a possibility of an independent determination of the
fine structure constants from the $g$ factor experiments with heavy ions.
For middle- and high-$Z$ ions the $g$ factor depends on $\alpha$ mainly 
via the relativistic binding correction in the Dirac formula (\ref{dir}).
It follows that the uncertainty of such a 
determination can be estimated by
\be
\frac{\delta \alpha}{\alpha} \sim \frac{1}{(\alpha Z)^2}
{\sqrt{(\delta g_{\rm exp})^2 +  (\delta g_{\rm th})^2} }\,,
\ee
where $\delta g_{\rm exp}$ and  $\delta g_{\rm th}$ are the uncertainties
of the experimental and theoretical
values of the $g$ factor, respectively. 
As one can see, at the given values of $\delta g_{\rm exp}$ and $\delta g_{\rm th}$
 the uncertainty of $\alpha$ decreases with $Z$ as $1/Z^2$.
A simple evaluation shows that
to get $\alpha$ to the same accuracy as it was obtained from the
free-electron $g$ factor experiment one should measure the $g$ factor
of H-like Pb to an uncertainty of about $3\times 10^{-10}$, provided the theoretical
value is known to the required precision. This uncertainty is by three
orders of magnitude bigger than the corresponding uncertainty in
the free-electron $g$ factor experiment.  This would seem very promising,
if we could calculate the theoretical value of the bound-electron $g$ factor
to the required accuracy. Unfortunately, the accuracy of the bound-electron
$g$ factor is strongly restricted by the nuclear size and polarization effects.
The use of the difference (\ref{dif1})  
of the $g$ factors of H- and Li-like ions,
which can be calculated to a very high accuracy, does not help since the $\alpha$-dependent 
term is significantly reduced in this difference. Instead,
 one should consider the corresponding
difference of the $g$ factors of H- and B-like ions of lead \cite{shabaev:2006:253002}: 
\begin{eqnarray} \label{g_dif}
g'=g_{(1s)^2 (2s)^2 2p_{1/2}}-\xi g_{1s}\,,
\end{eqnarray}
where $g_{(1s)^2(2s)^2 2p_{1/2}}$
and $g_{1s}$ are the $g$ factors of  $^{208}$Pb$^{77+}$
and   $^{208}$Pb$^{81+}$, respectively, and the parameter $\xi$
must be chosen to cancel the nuclear size effect. In Ref. \cite{shabaev:2006:253002}
it was shown that both the parameter $\xi$ and the difference $g'$
are very stable with respect to variations of the nuclear structure 
parameters. At the same time 
the $\alpha$-dependent term 
is reduced in this difference by about 4\% only. 
In Table \ref{g-dif} we give the uncertainties of $g'$ due 
to the uncertainty of the current value of $\alpha$ \cite{kinoshita:2014:1430003} and 
due to the nuclear polarization effect \cite{volotka:2014:023002}, which 
sets the ultimate accuracy limit up to which this difference can be
calculated. As one can see from the table, the uncertainty caused by the current value of
$\alpha$ is 
slightly bigger than the nuclear polarization limit. 
This means that the method has a potential to provide a determination of  { $\alpha$} to an 
accuracy  comparable to one obtained from the free-electron
$g$ factor study. 
 To achieve the required accuracy from the theoretical side, 
we need to calculate all two-loop and,
at least approximately, three-loop QED corrections. While the calculation
of the two-loop corrections seems quite realistic with the
most elaborated to-date methods \cite{yerokhin:2003:203,yerokhin:2009:R040501}, 
the evaluation of the three-loop
contributions requires development of essentially new methods.

\section{Conclusion}
In this paper we have reviewed the calculations of the various contributions to the $g$ factor
of highly charged ions. The experimental and theoretical investigations of the $g$ factors
 have already allowed the most stringent tests of one- and many-electron QED effects with
middle-$Z$ ions and provided the most precise determination of the electron mass.
The theoretical predictions for heavy ions suffer from the large uncertainties due 
to the nuclear structure effects. This 
prevents high-precision tests of bound-state QED on the two-loop level 
and restricts the possibilities for determinations of the fundamental constants by the direct comparison of the theory 
and experiment for heavy ions. 
It has been shown, however, that this restriction can be overcomed by studying the specific differences 
of the $g$ factors of H-, Li-, and B-like ions. The study of these differences
should allow tests of bound-state QED at strong fields and provide an independent
determination of the fine structure constant. The investigations of the isotope shifts of 
the $g$ factor with highly charged ions can give a unique opportunity to probe QED beyond the Furry picture.
They can also be used for a precise determination of the nuclear-charge-radius differences.
The study of the $g$ factors of highly charged ions with non-zero nuclear spin will
provide determinations of the nuclear magnetic moments with unprecedented accuracy.

\begin{table}
\caption{\label{H-like-Si} 
Theoretical contributions to the $g$  factor of H-like Si.}
\begin{tabular}{|l|l|}
\hline
Dirac value (point nucleus) & 1.993 023 571 6 \\ \hline
Free QED  & 0.002 319 304 4    \\ \hline
Binding QED  & 0.000 005 855 6(17) \\ \hline
Nuclear recoil  & 0.000 000 205 9 \\ \hline
Nuclear size & 0.000 000 020 5 \\ \hline
Total theory & 1.995 348 958 0(17) \\ \hline
Experiment \cite{sturm:2013:R030501} & 1.995 348 959 10(7)(7){(80)} \\ \hline
\end{tabular}
\end{table}

\begin{table}
\caption{\label{Li-like-Si} 
Theoretical contributions to the $g$  factor of Li-like Si.}
\begin{tabular}{|l|l|}
\hline
Dirac value (point nucleus) & 1.998 254 751 \\ \hline
Free QED  & 0.002 319 304     \\ \hline
Binding QED & 0.000 000 987(6)  \\ \hline
Interelectronic interaction & 0.000 314 809(6) \\ \hline
Nuclear recoil & 0.000 000 039(1) \\ \hline
Nuclear size & 0.000 000 003 \\ \hline
Total theory & 2.000 889 892(8) \\ \hline
Experiment \cite{wagner:2013:033003} & 2.000 889 890(2) \\ \hline
\end{tabular}
\end{table}

\begin{table}
\caption{\label{g-ca} 
Theoretical contributions to the isotope shift of the $g$  factor of H-like calcium:
 $^{40}{\rm Ca}^{19+} - \; ^{48}{\rm Ca}^{19+}$.}
\begin{tabular}{|l|l|}
\hline
Nuclear recoil: non-QED  $\sim m/M$   & $\;\;\; 0.000\, 000\, 048\, 657 $ \\ \hline
Nuclear recoil: non-QED  $\sim (m/M)^2$   & $ -0.000\, 000\, 000\, 026(2) $ \\ \hline
Nuclear recoil: QED  $\sim m/M$ &           $\;\;\; 0.000\, 000\, 000\,904$ \\ \hline
Nuclear recoil: QED  $\sim \alpha (m/M)$ &           $-0.000\, 000\, 000\, 038(3)$ \\ \hline
Nuclear size & $\;\;\; 0.000\, 000\, 000\, 032(78)$ \\ \hline
Total theory  & $ \;\;\; 0.000\, 000\, 049\, 529(78)$ \\ \hline
\end{tabular}
\end{table}

\begin{table}
\caption{\label{g-dif}
 The uncertainties of $g'$, defined by Eq. (\ref{g_dif}), for {Pb} due to
the current uncertainty of $\alpha$ and the nuclear polarization effect.}
\begin{tabular}{|l|l|l|}
\hline
Effect& $\delta g'$  $\;\;\;$  & $\delta g'/g'$ \hspace*{0.115cm}
\\ \hline
$1/\alpha=137.035999173(35)$
  &$0.4\times 10^{-10}$ $\;\;\;$ &$0.7\times 10^{-10}$  \hspace*{0.115cm}
\\ \hline
Nuclear polarization
  &$0.3\times 10^{-10}$  $\;\;\;$  &$0.5\times 10^{-10}$  \hspace*{0.115cm}
\\
\hline
\end{tabular}
\end{table}

%
%

%


\begin{acknowledgments}
This work was supported by RFBR (Grants No. 13-02-00630 and 14-02-31316), by 
SPbSU (Grants No. 11.38.269.2014 and No. 11.38.237.2015), by DFG, 
and by the FAIR--Russia Research Center.
\end{acknowledgments}


%

\end{document}